\documentclass{mn2e}
\usepackage{epsf}
\newcommand\msun   {M_{\odot}}

\title[The Faintest Accretors]
{The faintest accretors}

\author[A. R. King, R. Wijnands] {A. R. King$^{1, 2}$, R. Wijnands$^2$ \\
$^1$Theoretical Astrophysics Group, University of Leicester, Leicester
LE1 7RH\\ $^2$ Astronomical Institute ``Anton Pannekoek'' Kruislaan
403, University of Amsterdam, 1098 SJ Amsterdam, Netherlands}

\date{Accepted 2005 November 15. Received 2005 October 28; in original form 2005 August 1}

\volume{000}

\pagerange{\pageref{firstpage}--\pageref{lastpage}} \pubyear{2005}

\begin{document}

\label{firstpage}

\maketitle

\begin{abstract}
Recent X--ray observations have detected a class of very faint X--ray
transients in the Galaxy which probably reveal a previously
unrecognised type of accretion on to neutron stars or black holes.  We
show that these systems cannot have descended from binaries with
stellar--mass components of normal composition. Accretion of
hydrogen--depleted matter on to stellar--mass black holes can account
for individual systems, but requires that these transients should be
observed to repeat within a few years, and does not explain why the
class is distinctly faint.

Two other explanations appear to be quite natural. One invokes
accretion by neutron stars or stellar--mass black holes from
companions which were already brown dwarfs or planets when the systems
formed, i.e. which did not descend from low--mass stars. The other
possibility is that these systems are the endpoints of primordial
(zero--metallicity) binaries in which the primary was extremely
massive, and collapsed to a black hole of mass $\ga 1000\msun$. The
(primordial) companion must by now have reached an extremely low mass
($\la 0.01\msun$) and be transferring mass at a very low rate to the
black hole. This picture avoids the main difficulty encountered by
models invoking intermediate--mass black hole formation at
non--primordial metallicities, and is a natural consequence of some
current ideas about Population III star formation.

\end{abstract}


\section{Introduction}

Recent sensitive X--ray observations of Galactic fields (e.g., Sakano
et al.~2005, Muno et al.~2005a) have revealed the existence of a class
of very faint X-ray transients (henceforth VFXTs) with peak outburst
luminosities between $10^{34}$ and $10^{36}$~erg~s$^{-1}$, and
quiescent states at least a factor 10 below their peak values. A large
number of them are found very close (within 10 arcminutes; e.g., Muno
et al.~2005a) to Sgr A* although several are found at larger distances
(see, e.g., Hands et al.~2004 who reported on a VFXT about 20 degrees
away from Sgr A*).  Accretion on to a black hole or neutron star is
the most likely origin of this luminosity, as only one white dwarf
accretor (GK Per, Watson et al., 1985) is observed to have X--ray
outbursts as bright.  The outburst peak luminosities of the VFXTs are
several orders of magnitude lower than the peak luminosities observed
for the classic X--ray transients in our Galaxy (e.g., Chen et
al. 1997).  Observational limitations mean that our knowledge of the
duty cycle of the VFXTs is sketchy at best (Muno et
al.~2005a). However if this is $\la 10$~\%, as is common for the
brighter X--ray transients, typical accretion efficiencies
$10^{20}$~erg~g$^{-1}$ imply that the mean accretion rates on to these
black holes or neutron stars in VFXTs must be extraordinarily low,
i.e.  $\la 10^{-13}\msun~{\rm yr}^{-1}$. These correspond to mean
luminosities $\la 10^{33}$~erg~s$^{-1}$ and thus very cool accretion
discs. This means that the thermal--viscous disc instability can
operate, switching the accretion discs between high and low viscosity
states, and thus making the systems transient (King, Kolb \& Burderi,
1996; cf King, 2000). Hence despite their very low mean luminosities,
these systems are potentially detectable when using sensitive X-ray
instruments (e.g., {\it Chandra} or {\it XMM-Newton}) because they
save up their accretion energy for release in short outbursts.

The mass transfer rates of the VFXTs are a factor 10--100 less than
the mean rates in the {\em BeppoSAX} transients reported by Heise et
al.~(1998). King (2000) suggested that these {\em BeppoSAX} transients
might be low--mass X--ray binaries (LMXBs) which had passed the
minimum orbital period, and are currently accreting from very
low--mass ($\la 0.1\msun$) companions which are largely
degenerate. Detailed calculations by Bildsten \& Chakrabarty (2001)
tend to support this picture. The fainter VFXTs reported above clearly
present a challenge to theory, which this paper takes up.

\section{Models for very faint X-ray transients}

The source properties (e.g., spectral behaviour; Sakano et al.~2005;
temporal behaviour; Muno et al.~2005a, 2005b) indicate that the VFXTs
are probably not a homogeneous set of sources and several models may be
needed to explain all of them. One possibility is that these
transients are wind--fed X--ray transients: the accretor (apparently
always a strongly magnetic neutron star) accretes from the equatorial
disc of a Be star companion. The outbursts can be arbitrarily
faint. However no optical/IR companions of VFXTs have so far been
detected, suggesting that most VFXTs harbour companions fainter than
B2 IV stars (e.g., Muno et al.~2005a). This in turn suggests that at
least a sizable fraction of these systems accrete from relatively
low--mass companions. Moreover, two VFXTs (GRS 1741.9--2853; Cocchi et
al., 1999\footnote{We note that GRS 1741.9--2853 has shown peak
outburst X--ray luminosities of a few times $10^{36}$ erg s$^{-1}$,
which would place it just outside our maximum X-ray luminosity of
$10^{36}$ erg s$^{-1}$ for classification as VFXTs.}  and SAX
J1828.5--1037; Cornelisse et al.~2002; Hands et al.~2004) have shown
type I X--ray bursts. Only neutron stars accreting from low--mass
companions have so far been observed to show such bursts, providing
additional evidence that a significant fraction of the VFXTs might be
LMXBs. However as we shall see, the requirement that mass transfer
from a Roche--lobe-filling companion should drop to values $\la
10^{-13}\msun~{\rm yr}^{-1}$ within a Hubble time poses
extremely tight constraints on the nature of these
systems. Accordingly we briefly consider other types of explanations
for the VFXTs.

A number of known LMXBs appear fainter than their true luminosities
because we view them at unfavourable angles (e.g. White \& Holt,
1982), and at least one VFXT appears fainter than its intrinsic
luminosity because of inclination effects (Muno et al.~2005b). One
might ask if the VFXT class are the `accretion disc corona', and
`dipping' sources belonging to an intrinsically faint population of
LMXBs. However the range of inclination angles over which ADC and
dipping behaviour can be seen is quite small. If we generously allow a
range from $i = 60^{\circ}$ to $i = 90^{\circ}$ for this behaviour, a
random distribution of orbital inclinations should give roughly equal
numbers of brighter LMXBs mixed in with the VFXTs (solid angle
$\propto \cos i)$.  This is quite contrary to the observed
distribution around the Galactic Centre, where observations go
deepest, which is dominated by VFXTs (e.g., Muno et al. 2005a;
Wijnands et al. 2005).

Another idea invoking orthodox LMXB evolution is that of long--term
variability. Binaries transferring mass via Roche--lobe overflow can
deviate significantly from the evolutionary mean mass transfer rate
for a timescale of order $(H/R_2)t_M$, where $H/R_2 \simeq 10^{-4}$
(Ritter, 1988) is the ratio of atmospheric scale-height $H$ to radius
$R_2$ for the mass--losing star, and $t_M = -M_2/\dot M_2$ is the mass
transfer timescale ($M_2 =$ companion mass, $-\dot M_2 = $ mass
transfer rate). This can happen for a variety of reasons, e.g. the
motion of starspots under the $L_1$ point (cf King \& Cannizzo, 1998;
King, 2006), or through irradiation--driven cycles (cf King et al., 1995).

For the VFXTs the mass transfer rate can thus depart
from the mean for a timescale $\sim 10^9 m_2$~yr, where $m_2 =
M_2/\msun \la 1$ for an LMXB. This is significantly less than the age
of the Galaxy and (by construction) significantly less than the
lifetime of the LMXB. Thus again the VFXTs should be accompanied by a
set of much brighter systems at or above the evolutionary mean mass
transfer rate, and indeed at intermediate luminosities. While it is
impossible to rule this possibility out completely, one would need a
candidate mechanism (say of radius variations of the companion) which
would cause the transfer rate to decrease by very large
factors. Precisely the same argument applies to systems just beginning
Roche lobe overflow.

As explained above, no early type companion has so far been detected
for any VFXTs, suggesting that most are not wind accreting systems.
One might seek to evade this difficulty by invoking accretion from the
wind of a fainter low--mass star. However, the system must eventually
evolve into a state where the companion fills its Roche lobe, giving a
brighter and probably longer--lasting system. Again the lack of
brighter systems mixed in with the VFXTs makes this idea implausible.

{\it INTEGRAL} has detected a class of X--ray binaries with such high
intrinsic absorption (equivalent hydrogen columns $\ga
10^{23-24}$~cm$^{-1}$) that much of the primary X--ray emission is
absorbed and reradiated at longer wavelengths (Walter et al.,
2003). However, most of these systems appear to be high--mass systems,
with the wind of the companion providing the absorption. As discussed
above, such systems are not good models for VFXTs.

We conclude that none of these alternative ideas provides a tenable
picture of the VFXTs. However, all of the arguments against them
invoke the properties of the whole population, rather than individual
source characteristics. This means that we cannot rule out the
possibility that {\em some} of the VFXTs belong to one or other of
these classes. Indeed this is extremely likely.

\section{Accretion from very low--mass companions}

Here we test whether Roche lobe overflow from a low--mass degenerate
companion offers a viable explanation for VFXTs. For a simple picture
we model the companion as an $n = 3/2$ polytrope with radius
\begin{equation}
R_2 \simeq 10^9(1+X)^{5/3}m_2^{-1/3}~{\rm cm}
\label{r2}
\end{equation}
where $m_2$ is the companion mass in $\msun$. This permits an analytic
description of the evolution and thus makes the dependences on various
parameters transparent. Specifically we use the formulae from King
(1988), but retain explicitly the dependences on primary mass $m_1$
(in $\msun$) and fractional hydrogen content by mass
$X$. We consider the effect of relaxing these assumptions at the end
of this Section.

Gravitational radiation drives a mass transfer rate
\begin{equation}
-\dot m_2 = 1.3\times 10^{-3}(1 + X)^{-20/3}m_1^{2/3}m_2^{14/3}~{\rm yr}^{-1}
\label{mdot}
\end{equation}
Integrating this gives
\begin{equation}
m_2 = m_0[1 + 4.8\times 10^{-3}(1+X)^{-20/3}m_1^{2/3}m_0^{11/3}t]^{-3/11}
\end{equation}
where $m_0$ (in $\msun$) is the companion mass at time $t=0$.  
Then (\ref{mdot}) gives the transfer rate as an explicit function of time:
\begin{equation}
-\dot m_2 = {1.3\times 10^{-3}(1+X)^{-20/3}m_1^{2/3}m_0^{14/3}\over[1 +
 4.8\times 10^{-3}(1+X)^{-20/3}m_1^{2/3}m_0^{11/3}t]^{14/11}}.
\label{mdot2}
\end{equation}

We see from \ref{mdot} that to ensure the required transfer rates $\la
10^{-13}\msun {\rm yr}^{-1}$ to explain the VFXTs, current masses $m_2
\la 0.014m_1^{-1/7}$ are needed for hydrogen--rich companions, and
still smaller ones ($m_2 \la 0.0068m_1^{-1/7}$) for hydrogen--poor
companions. Clearly one way of ensuring these would be to postulate
that the companions came into contact with masses of these
orders. This would require brown--dwarf or planetary companions. If we
assume instead that the companion came into contact with a normal
stellar mass, but has been whittled down to the required low current
mass through mass transfer in the past, we must take $m_0 \ga
0.1$. Then for times $t >> t_0 = 200(1+X)^{20/3}m_1^{-2/3}m_0^{-11/3}$
the constant 1 in the denominator is negligible: since $t_0 \la
2\times 10^8$~yr (even taking $X = 1, m_0 = 0.1$) this is essentially
all of the evolution. Then we get the simplified formula
\begin{equation}
-\dot m_2 = 2\times
 10^{-13}(1+X)^{20/11}m_1^{-2/11}(t_{10})^{-14/11}~\msun~{\rm yr}^{-1}
\label{mdot3}
\end{equation}
where $t_{10} =  t/10^{10}~{\rm yr}$: we note that estimates of the
Hubble time require $t_{10} \la 1.37$.

The result (\ref{mdot3}) shows that with normal primary masses ($m_1
\la 10$), hydrogen--rich companions are ruled out, and even
hydrogen--poor companions ($X = 0$) are not promising candidates for
explaining VFXTs. The general problem is that the predicted mass
transfer rates are too low to reduce the companion from a normal
stellar mass to the very small values $\la 0.01\msun$ needed to
explain the VFXTs within the age of the Galaxy. This problem is
exacerbated if we consider more detailed models for the binary
evolution. Bildsten \& Chakrabarty (2001) show that the companion is
likely to evolve at constant radius rather than the cold $n = -1/3$
polytrope considered above. This makes the time to reach the required
low masses still longer. We conclude that stretching `standard' LMXB
evolution to its limits does not explain the VFXTs.

\section{Discussion}

We have seen above that standard LMXB evolution is too slow for
initially stellar--mass companions to reach the very low masses that
the VFXTs probably have. There are evidently three ways out of this
impasse. Either

(a) VFXT companions are hydrogen--poor, or

(b) VFXTs are born with very low companion masses, or

(c) VFXTs evolve more quickly than assumed above, i.e. they have
larger accretor masses $m_1$ than assumed there.

\subsection{(a) Hydrogen--poor companions}

From (\ref{mdot3}) we see that transfer rates $\sim 10^{-13}\msun~{\rm
 yr}^{-1}$ are just possible if $X=0, t_{10} = 1.37$ and $m_1 >
 5$. Thus VFXTs would have to be extremely old black--hole LMXBs with
 hydrogen--poor companions. This explanation requires that the
 transfer rates are not significantly lower than the $10^{-13}\msun~{\rm
 yr}^{-1}$ inferred by assuming a 10\% duty cycle for the transient
 outbursts, and hence that VFXT outbursts are seen to repeat in a few
 years. In addition, any optical identifications should show an
 absence of hydrogen. 

Before accepting this idea as a viable model for VFXTs it is clear
that more elaborate evolutionary calculations with full stellar models
for the companion are needed. In any case, the main weakness of this
fairly conservative explanation is that it gives no reason for the
defining feature of the VFXT class, namely that their inferred
luminosities are distinctly lower than other accreting systems.

\subsection{(b) Low initial companion masses}

LMXBs with very low mass companions are known, e.g. HETE J1900.1-2455
(Kaaret et al., 2005). Here the reported companion mass could be as
low as $0.016\msun$, i.e. almost down at the value inferred above for
most VLXTs. Unless HETE J1900.1-2455 is in an unusual shortlived stage
however, the work of this paper shows that the companion mass must
have been similarly low {\it at formation}, and not the result of mass
transfer from an initially stellar--mass companion. The same
conclusion holds if we want to explain a significant fraction of the
VFXT population this way. Evidently the companion must start life as a
brown dwarf, or possibly a planet, and somehow survive the eventful
evolution of its stellar companion (expansion, and possible
supernova). A further constraint is that this object must eventually
fill its Roche lobe, despite having little orbital energy to
contribute to any common--envelope ejection. Despite these
difficulties, the mere existence of HETE J1900.1-2455 shows that this
type of model for the VFXTs is clearly possible. To see if this model
explains why VFXTs have distinctly low luminosities probably requires
one to understand the initial masses of brown dwarfs.

\subsection{(c) Larger--mass accretor}

Equation (\ref{mdot3}) shows that hydrogen--rich systems which
initially have stellar--mass companions require accretor masses
\begin{equation}
m_1 \ga 5(1+X)^{10} \simeq 1000
\label{imbh}
\end{equation}
Hence this type of explanation requires VFXTs to contain
intermediate--mass black holes (IMBHs). There are claims that such
black holes may form by rapid dynamical processes in dense clusters
(e.g., Portegies Zwart et al. 2004; G\"urkan et al, 2004) or through
capture of the nuclei of small satellite galaxies (e.g., King \&
Dehnen, 2005). However the fact that the VFXTs are evidently extremely
old systems offers an attractive alternative. 

Namely, this kind of VFXT is the natural endpoint for a {\it
primordial} (Population III) binary containing a very massive star
with a normal--mass companion. At primordial metallicities (say $Z <
10^{-4}$) the formation of very high--mass stars is quite possible:
moreover, the lack of metals means that they retain most of their mass
on collapsing to form a black hole, i.e. they form an IMBH (cf Madau
\& Rees, 2001). (IMBH formation scenarios invoking formation in
clusters fail this test, as at non--primordial metallicities massive
stars lose almost all their mass via metal--driven winds and
ultimately form only stellar--mass black holes.) As we have seen, the
evolution of such binaries effectively grinds to a halt once the
companion has reached masses $m_2 \sim 10^{-2}$. Such systems would
appear as VFXTs. A second potentially attractive feature also appears
naturally. Such systems would probably spiral toward the Galactic
Centre by dynamical friction. This offers a possible explanation if
the reported excess of VFXTs near the Galactic Centre is confirmed
(Muno et al.~2005a).

Observing programmes to monitor the VFXTs in the Galactic Centre
region are continuing (see Wijnands et al.~2005 for details). We have
seen that extremely low mass transfer rates in X--ray sources can pose
very tight constraints and suggest far--reaching conclusions. So for
once it may be that {\it non}--detections in repeated observations,
which drive down the mean luminosities of these objects, could be as
interesting as detections.

\section{Acknowledgments} 

ARK acknowledges a Royal Society--Wolfson Research Merit Award, and
the hospitality of the Astronomical Institute ``Anton Pannekoek'',
University of Amsterdam, where this paper was written.

\label{lastpage}


\begin{thebibliography}{}

\bibitem{}
Bildsten, L., Chakrabarty, D., 2001, ApJ, 557, 292

\bibitem{}
Chen, W., Shrader, C.R., Livio, M., 1997, ApJ, 491, 312

\bibitem{} 
Cocchi, M., Bazzano, A., Natalucci, L., Ubertini, P., Heise, J., Muller, J.M.,
in 't Zand. J.J.M., 1999, A\&A 346, L45

\bibitem{}
Cornelisse, R., Verbunt, F., In 't Zand, J.J.M., Kuulkers, E., Heise,
J., Remillard, R.A., Cocchi, M., Natalucci, L., Bazzano, A., Ubertini,
P., 2002, A\&A, 392, 885

\bibitem{}
G\"urkan, M.A., Freitag, M., Rasio, F.A., 2004, ApJ, 604, 632

\bibitem{}
Hands, A.D.P., Warwick, R.S., Watson, M.G., Helfand, D.J., 2004,
MNRAS, 351, 31

\bibitem{} 
Haswell, C.A., King, A.R., Murray, J.R., Charles, P.A., 2001, MNRAS,
321, 475

\bibitem{}
Heise, J., in 't Zand, J.J.M., Smith, S.M.J., Muller, M.J., Ubertini,
P., Bazzano, A., Cocchi, M., Natalucci, L., 1999, ApL\&C, 38, 297

\bibitem{}
Kaaret, P., Morgan, E.H., Vanderspek, R., Tomsick, J.A., 2005, ApJ, in
press (astro--ph/0510483)

\bibitem{}
King, A.R., 1997, in {\it Relativistic Gravitation and Gravitational
  Radiation} eds. J.--A. Marck \& J.--P. Lasota, Cambridge University
press, Cambridge, p. 105

\bibitem{}
King, A.R., 2000, MNRAS, 315, L33

\bibitem{} 
King, A.R., Cannizzo, J.K., 1998, ApJ, 499, 348

\bibitem{}
King, A.R., Dehnen, W., 2005, MNRAS, 357, 275 

\bibitem{}
King, A.R., Frank, J., Kolb, U., Ritter, H., 1995, ApJ 444, L37

\bibitem{} 
Madau, P., Rees, M.J., 2001, ApJ 551, L27

\bibitem{}
Muno, M.P., Baganoff, F.K., Arabadijs, J.S., 2003, ApJ, 598, 474

\bibitem{}
Muno, M.P., Pfahl, E., Baganoff, F.K., Brandt, W.N., Ghez, A., Lu, J.,
Moris, M.R., 2005a, ApJ, 622, L113

\bibitem{}
Muno, M.P., Lu, J.R., Baganoff, F.K., Brandt, W.N., Garmire, G.P.,
Ghez, A.M., Hornstein, S.D., Morris, M.R. 2005b, ApJ, submitted
(astro-ph/0503572)

\bibitem{} 
Portegies Zwart, S.F., Baumgardt, H., Hut, P., Makino, J.,
McMillan, S.L.W., 2004, Nat., 428, 724

\bibitem{}
Ritter, H., 1988, A\&A, 202, 93

\bibitem{}
Sakano, M., Warwick, R.S., Decourchelle, A., Wang, Q.D., 2005, MNRAS,
357, 1211

\bibitem{} 
Walter, R., Rodriguez, J., Foschini, L., de Plaa, J.,
Corbel, S., Courvoisier, T. J.-L., den Hartog, P. R., Lebrun, F.,
Parmar, A. N., Tomsick, J. A., Ubertini, P., 2003, A\&A, 411, 427

\bibitem{}
Watson, M.G., King, A.R., Osborne, J., 1985, MNRAS, 212, 917

\bibitem{}
White, N.E., Holt, S.S., 1982, ApJ, 257, 318

\bibitem{}
Wijnands, R., In 't Zand, J.J.M., Rupen, M., Maccarone, T., homan, J.,
Cornelisse, R., Fender, R., Grindlay, J., van der Klis, M., Kuulkers,
E., Markwardt, C.B., Miller-Jones, J.C.A., Wang, Q.D., 2005, A\&A,
submitted (astro-ph/0508648)


\end{thebibliography}
\end{document}